\documentclass[superscriptaddress, nofootinbib, twocolumn, amsmath,amssymb, aps, pra, notitlepage, longbibliography]{revtex4-1}

\usepackage{dcolumn}
\usepackage{bm}
\usepackage{epsfig}
\usepackage{graphicx}
\usepackage{latexsym}
\usepackage{amsfonts}
\usepackage{setspace}
\usepackage{graphicx}
\usepackage{amsmath}
\usepackage{verbatim}
\usepackage{color}
\usepackage{hyperref}
\usepackage{nccmath}
\usepackage{upgreek}
\usepackage{siunitx}

\begin{document}

\title{Brillouin spectroscopy of a hybrid silicon-chalcogenide waveguide with geometrical variations}

\author{Atiyeh Zarifi$^{1,2,\ast}$, Birgit Stiller$^{1,2}$, Moritz Merklein$^{1,2}$, Yang Liu$^{1,2}$, Blair Morrison$^{1,2}$, Alvaro Casas-Bedoya$^{1,2}$, Gang Ren$^{3}$, Thach G. Nguyen$^{3}$, Khu Vu$^{4}$, Duk-Yong Choi$^{4}$, Arnan Mitchell$^{3}$, Stephen J. Madden$^{4}$ and Benjamin J. Eggleton$^{1,2}$\\
\small{ \textcolor{white}{blanc\\}
$^{1}$Sydney Nano Institute (Sydney Nano), The University of Sydney, NSW 2006, Australia.\\
$^{2}$Institute of Photonics and Optical Science (IPOS), School of Physics, The University of Sydney, NSW 2006, Australia.\\
$^{3}$School of Engineering, RMIT University, Melbourne, VIC 3001, Australia.\\
$^{4}$Laser Physics Centre, Research School of Physics and Engineering, Australian National University, Canberra, ACT 2601, Australia.\\
$^{\ast}$atiyeh.zarifi@sydney.edu.au}}


\begin{abstract}
Recent advances in design and fabrication of photonic-phononic waveguides have enabled stimulated Brillouin scattering (SBS) in silicon-based platforms, such as under-etched silicon waveguides and hybrid waveguides.
Due to the sophisticated design and more importantly high sensitivity of the Brillouin resonances to geometrical variations in micro- and nano-scale structures, it is necessary to have access to the localized opto-acoustic response along those waveguides to monitor their uniformity and maximize their interaction strength. 
In this work, we design and fabricate photonic-phononic waveguides with a deliberate width variation on a hybrid silicon-chalcogenide photonic chip and confirm the effect of the geometrical variation on the localized Brillouin response using a distributed Brillouin measurement.

\end{abstract}

\maketitle



Stimulated Brillouin scattering (SBS) is an inelastic scattering process in which energy of the optical pump wave is coupled to a frequency down-shifted Stokes wave through a moving acoustic wave. SBS enables a range of applications such as microwave signal processing\cite{Marpaung2013,Pagani2014a,Santagiustina2013}, microwave signal generation \cite{Li2013,Merklein2016}, light storage \cite{Merklein2017,Dong2015,Galland2014,Fang2016b} and sensing \cite{Hotate2012,Song2006,Antman2016,Jia2017,Godet2017,Nikles1996a,Bao1993a}. There has been a strong interest to activate SBS-enabled functionalities in silicon platforms. This is mainly due to the fact that silicon photonics is capable of integrating multiple functions such as modulators and detectors into a single chip using the same facilities as for microelectronics circuits \cite{Chrostowski2013}.
However, harnessing SBS in silicon-based platforms is challenging. This difficulty is mainly attributed to the acoustic mode leakage in silicon on insulator (SOI) devices which results in a poor opto-acoustic overlap \cite{Poulton2013}. Moreover, nonlinear loss caused by two photon absorption (TPA) and free carrier absorption (FCA) limits the coupled pump power and hence the nonlinear coupling to acoustic waves \cite{Kittlaus2015}. 
A number of approaches have been demonstrated to harness SBS in silicon, such as under-etched silicon structures, which allows for strong opto-acoustic overlap in SOI platforms \cite{VanLaer2015} and suspended silicon membranes which allows for independent photonic and phononic design and therefore significantly lower propagation loss \cite{Kittlaus2015}. 

\begin{figure}[b!]
\centering
\includegraphics[width=\linewidth]{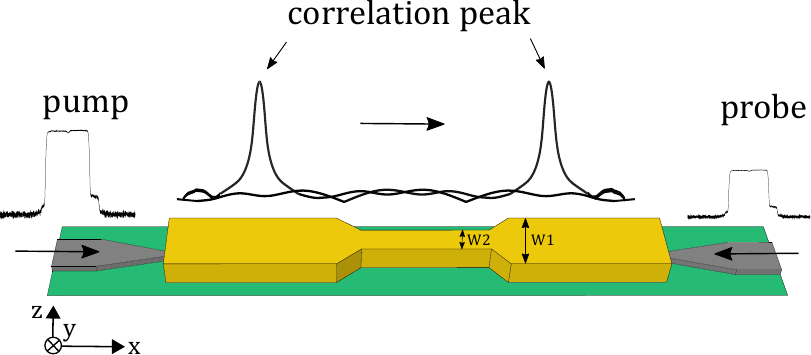}
\caption{Schematic of the BOCDA scheme in a hybrid waveguide. The gray region shows the silicon grating couplers tapers into the chalcogenide waveguide which is colored in yellow. $w_{1}$ and $w_{2}$ are 1.9\,$\upmu m$ and 1.08\,$\upmu m$, respectively. The waveguide is scanned by changing the position of the correlation peak along the waveguide.}
\label{fig:test_concept}
\end{figure}

Recently, the hybrid integration of soft glasses on a silicon-based platform has been demonstrated, in which SBS interaction takes place in the soft glass while the coupling into and out of the waveguide is through silicon grating couplers \cite{Morrison2017}. This approach takes advantage of a large SBS coefficient and negligible nonlinear loss in the soft glass ($As_{2}S_{3}$) while giving access to the rich library of functional silicon devices, which is promising for bringing SBS-functionalities into a silicon integrated photonics platform. 
Such novel design requires a characterization system to evaluate the opto-acoustic interaction at critical design points such as bends and tapers. This is of critical importance since structural variations along the waveguide results in Brillouin resonance broadening and reduces the Brillouin scattering efficiency \cite{Wolff2016}. Therefore, an SBS-based distributed measurement technique with high spatial resolution is required to map SBS response against the position in the few-mm long hybrid waveguides.

\begin{figure}[t!]
\centering
\includegraphics[width=\linewidth]{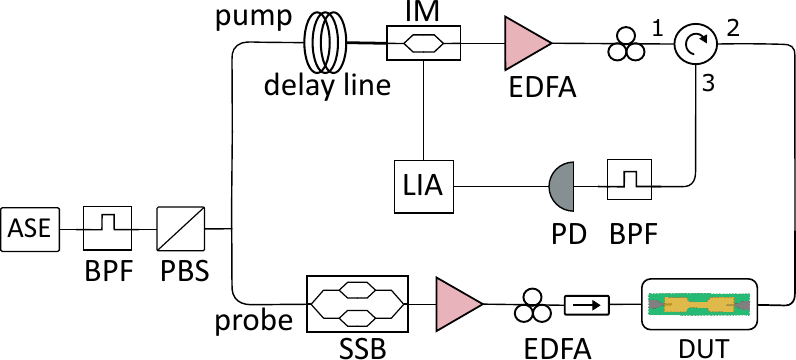}
\caption{Schematic of the BOCDA setup. BPF: band pass filter, PBS: polarization beam splitter, SSB: single side-band modulator, IM: intensity modulator, LIA: lock-in amplifier, PD: photo-detector, EDFA: Erbium doped fiber amplifier, DUT: device under test.}
\label{fig:setup_1}
\end{figure}

Brillouin scattering has been widely employed to study the uniformity and acoustic properties of optical fibers and micro-structures. Distributed Brillouin scattering has been employed to map the excited surface acoustic waves in microfibers \cite{Chow2018} and to study the effect of diameter and microstructure fluctuations in photonics crystal fibers \cite{Stiller2010,B.Stiller2012,Beugnot2007}. In-situ monitoring of a tapered fiber diameter based on backward Brillouin spectroscopy was reported by \cite{Godet2017}. On-chip waveguide characterization was also performed using distributed SBS measurement on a planar lightwave circuit (PLC) \cite{Hotate2012} and more recently on a chacogenide photonic waveguide \cite{Zarifi2017}. Forward Brillouin scattering, also called guided acoustic wave Brillouin scattering (GAWBS) was used to characterize the diameter of a tapered fiber \cite{Jarschel2018,Florez2016a,Kang2008} and the core diameter of photonics crystal fibers \cite{Zhong2015a}. 

This work presents a custom-designed photonic waveguide for distributed sensing in order to confirm geometric variations of the photonic waveguides with a confirmed spatial resolution. Brillouin optical correlation domain analysis (BOCDA) technique based on an amplified spontaneous emission (ASE) of an Erbium doped fiber \cite{Zarifi2017,Cohen2014a} was employed to characterize the silicon-chalcogenide hybrid waveguide. The hybrid integration approach underpinned by the flexible electron beam lithography (EBL) patterning technique enables us to introduce controlled variations in the waveguide in order to demonstrate and confirm the spatial resolution and the sensing capability of the distributed SBS measurement system.



A schematic of the hybrid waveguide is shown in Fig. \ref{fig:test_concept}. The design starts with a SOI platform consisting of silicon grating couplers to couple light into the standard silicon nanowire ($220\times450$\,nm) which continues for 2\,mm. The width of the silicon nanowire is adiabatically reduced to 150\,nm before it interfaces with the 690\,nm thick chalcogenide ($As_{2}S_{3}$) region so that the optical mode can progressively transfers from the silicon to the chalcogenide waveguide. The chalcogenide strip waveguide is 1.9\,$\upmu$m wide which is adiabatically reduced to 1.08\,$\upmu$m in the middle for 2\,mm. The total length of the chalcogenide waveguide is 6\,mm and the length of the tapers is 15\,$\upmu$m.

\begin{figure}[t!]
\centering
\includegraphics[width=\linewidth]{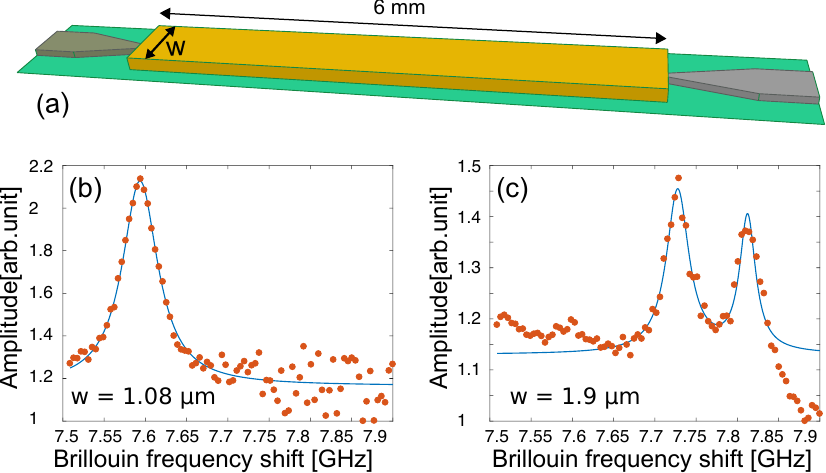}
\caption{a) Schematic of the reference silicon-chalcogenide hybrid waveguide with constant width (w). b) Integrated SBS response in 1.08$\upmu$m-wide waveguide, and c) 1.9$\upmu$m-wide waveguide. (The dots show the measured data and the solid blue line shows the Lorentz fit in the plots).}
\label{fig:figure_3}
\end{figure}

Since the Brillouin frequency shift (BFS) is sensitive to the waveguide cross section, the variation in the width can be detected by monitoring the BFS along the waveguide.
In the back-scattered SBS process the coupling between the optical pump and the Stokes waves through the traveling acoustic wave is most efficient when the three waves are phase-matched. That is, if the propagation constant of the pump and Stokes waves are $k_{p}$ and $k_{s}$, then the propagation constant of the acoustic wave must be $q = k_{p}-k_{s}$ \cite{Eggleton2013}. Under this condition, one or more acoustic modes will exist for a given optical mode, whose overlap with the optical mode creates the Brillouin gain profile and determines the BFS. The BFS is defined by the following equation:

\begin{equation}
\Omega _{B}=\frac{2n_\mathrm{eff}\mathrm{v}_{a}}{\lambda }, 
\label{eq:delta nu}
\end{equation}
where $\mathrm{v} _{a}$ is the velocity of each acoustic mode in the medium, $n_\mathrm{eff}$ is the effective refractive index and $\lambda$ is the pump wavelength. As a result, any change in waveguide geometry affects the effective refractive index and the acoustic mode resonances \cite{Chow2018} and consequently the BFS.


\begin{figure*}[t]    
\centering
\includegraphics[width=0.8\linewidth]{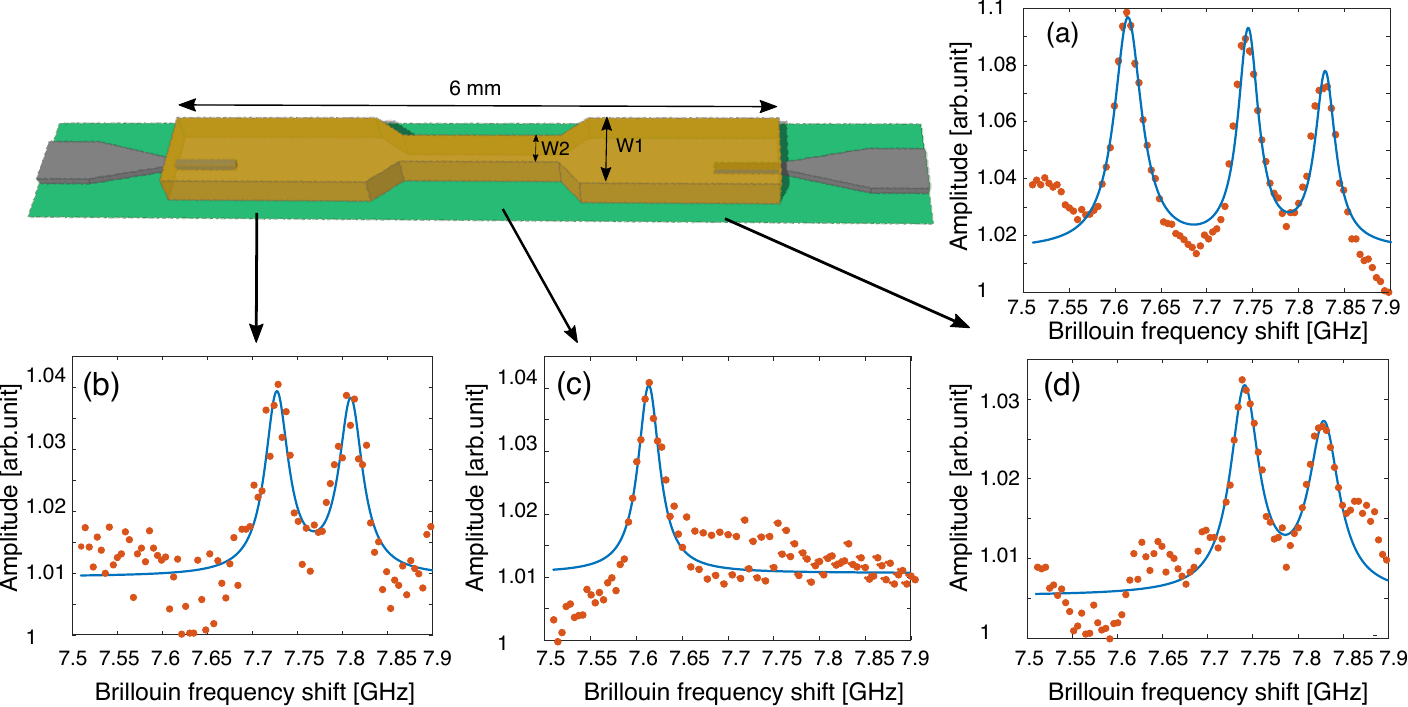}
\caption{a) BOCDA measurement of the waveguide with 12 mm spatial resolution. BOCDA measurement with 2\,mm spatial resolution in  b) $w_{1} = 1.9 \upmu m$ wide region at the left,  c) $ w_{2} = 1.08 \upmu m$ wide region in the middle, and  d) $ w_{1} = 1.9 \upmu m$ wide region at the right side of the wavegudie. (The dots show the measured data and the solid blue line shows the Lorentz fit in all the plots).}
\label{fig:experiment}
\end{figure*}

The distributed measurement of SBS responses in the waveguide with 2 mm spatial resolution is enabled by the BOCDA technique \cite{Song2006}. In this method, the pump and the probe signals are driven from a random noise source (in this case a filtered ASE source) in such a way that the efficient SBS interaction between the two waves occurs only at the correlation peak. As illustrated in the schematic in Fig. \ref{fig:test_concept}, the local SBS responses along the waveguide could be obtained by moving the position of the correlation peak. More details about this technique is provided in references \cite{Cohen2014a,Zarifi2017}.

Unlike the time-dependent techniques, where the spatial resolution is limited by the pulse width \cite{Fellay1997}, the spatial resolution in this method is mainly limited by the stochastic nature of the ASE source which limits the signal to noise ratio (SNR) of the measurement. The spatial resolution in the ASE-based BOCDA measurement is given by $\frac{1}{2}V_{g}\Delta t$, where $V_{g}$ is the group velocity of the optical mode and $\Delta t$ is inversely related to the spectral bandwidth of the pump and the probe.

A schematic of the experimental setup is shown in Fig. \ref{fig:setup_1}, where an ASE source is sent to a band pass filter and after passing through a polarization beam splitter, is split to the pump and probe signals.
In the probe arm, an RF frequency equal to the BFS is applied to the modulator to implement single side-band carrier suppressed (SSB-CS) modulation, generating the probe by downshifting the pump. By sweeping the probe signal using an RF synthesizer, the Brillouin gain spectrum around the BFS is measured. An intensity modulator in the pump arm is used to generate 500\,ns pump pulses with repetition rate of 10\,$\upmu$s. The same RF pulse generator that is used in the pump arm, triggers the lock-in amplifier (LIA) at the detection stage. The use of an LIA in the setup improves the SNR.

The distributed measurement along the waveguide is performed by changing the relative delay between the pump and the probe arms using a delay line. The spectral resolution of the distributed measurements is 4.5\,MHz. 
After collecting the backscattered signal at port 3 of the circulator, a sharp optical filter is used before the photo-detector in order to remove the pump back reflection from the grating couplers as much as possible. The filtering of the pump back reflection in this experiment is necessary since the pump back-reflection from the silicon grating coupler is significant. However, filtering at this stage is challenging since the back-reflected pump and the amplified probe have a broad spectral overlap. Therefore, it is not possible to filter the pump back reflection without cutting part of the SBS signal. 



To confirm the BFS associated with each waveguide geometry, we first measured the integrated SBS response in two hybrid reference waveguides with \SI{1.08}{\um} and \SI{1.9}{\um} widths as shown in Fig. \ref{fig:figure_3}(a). The length of each reference waveguide is \SI{6}{\mm} and the width is constant along the waveguide. Figure. \ref{fig:figure_3}(b) and (c) show the integrated SBS response for \SI{1.08}{\um} wide and \SI{1.9}{\um} wide waveguides, respectively. As illustrated in Fig. \ref{fig:figure_3}, the BFS for the two reference waveguides with different cross sections happens at different frequencies as expected from Eq. (\ref{eq:delta nu}). In addition, the Brillouin spectrum of the wider reference waveguides consists of two BFS peaks, while the Brillouin spectrum of the narrower waveguide has only one BFS.
This is most likely due to the alignment mismatch between the silicon taper and the chalcogenide waveguide resulting in the excitation of some higher order optical modes in the wide waveguide. This is also further supported by the simulation result in Ref. \cite{Morrison2017}, suggesting that the chalcogenide waveguide with \SI{1.9}{\um} can accommodate up to three higher order TE modes, although some of them are very lossy.

We then changed the device under test to the hybrid waveguide with varying width and performed an integrated SBS measurement to observe the Brillouin spectrum of the entire waveguide. In order to perform this measurement, we set the spatial resolution of the measurement to twice the length of the waveguide (\SI{12}{\mm}) by filtering the ASE bandwidth to \SI{5}{\GHz}. The delay line is adjusted in such a way that the pump and the probe arms have exactly the same length so that the correlation peak occurs in the middle of the waveguide and covers the entire structure. Fig. \ref{fig:experiment}(a) illustrates the integrated SBS response of the hybrid waveguide. As seen in Fig. \ref{fig:experiment}(a), the Brillouin spectrum consists of three BFS at distinct frequencies of 7.60, 7.74 and 7.82 \SI{}{\GHz}.

 A higher spatial resolution is required to individually detect BFS peaks at different positions along the waveguide. Therefore, the ASE bandwidth was set to \SI{30}{\GHz} corresponding to \SI{2}{\mm} spatial resolution in the waveguide. A distributed measurement was performed by stepping the delay line with \SI{2}{\mm} steps along the waveguide. The local SBS response at different points of the hybrid waveguide are plotted in Fig. \ref{fig:experiment}(b) to (d). The Brillouin spectrum at \SI{1.9}{\um} wide section consists of two BFS at 7.74 and 7.82 \SI{}{\GHz}. However, in the middle section with \SI{1.08}{\um} width only one BFS peak is observed at \SI{7.60}{\GHz}. The Brillouin spectrum collected from different sections of the waveguide agree with the integrated measurement taken at reference waveguides with similar widths. 
 As seen in Fig. \ref{fig:experiment}(b) to (d), \SI{800}{\nm} change in the waveguide width results in \SI{140}{\MHz} change in the BFS for the fundamental mode, in agreement with Eq. (\ref{eq:delta nu}). However, since the linewidth of the measurement is in the order of \SI{40}{\MHz}, width variation as small as \SI{100}{\nm}(corresponding to \SI{20}{\MHz} BFS change) is expected to be detected by this system.
 The spatial resolution of the system is confirmed to be \SI{2}{\mm} since the section in the middle is clearly resolved without detecting the Brillouin response from the other sections of the waveguide.



In conclusion, we reported a distributed SBS measurement based on the BOCDA technique on a silicon-chalcogenide hybrid waveguide. Spatial resolution of \SI{2}{\mm} has been demonstrated by resolving a \SI{2}{\mm} feature on the waveguide. This measurement is a proof-of-principle for waveguide characterization in a silicon-based platform with an active SBS gain medium. In addition, this approach provides valuable information about geometry-dependent opto-acoustic responses for further design and fabrication improvement. 
The spatial resolution of this measurement was mainly limited by the strong pump back reflection from the grating couplers. Optical filtering of the pump back reflection was also limited due to the wide spectral overlap between the pump back reflection and the amplified probe for ASE bandwidth beyond 30 GHz. Further improvement in spatial resolution is therefore possible by reducing the back reflection from the grating couplers in the fabrication process.

\section*{Funding Information}

The Australian Research Council (ARC) Laureate Fellowship (FL120100029) and the Centre of Excellence program (CUDOS CE110001018).



\bibliography{OFS_Ref}

\begin{thebibliography}{34}%
\makeatletter
\providecommand \@ifxundefined [1]{%
 \@ifx{#1\undefined}
}%
\providecommand \@ifnum [1]{%
 \ifnum #1\expandafter \@firstoftwo
 \else \expandafter \@secondoftwo
 \fi
}%
\providecommand \@ifx [1]{%
 \ifx #1\expandafter \@firstoftwo
 \else \expandafter \@secondoftwo
 \fi
}%
\providecommand \natexlab [1]{#1}%
\providecommand \enquote  [1]{``#1''}%
\providecommand \bibnamefont  [1]{#1}%
\providecommand \bibfnamefont [1]{#1}%
\providecommand \citenamefont [1]{#1}%
\providecommand \href@noop [0]{\@secondoftwo}%
\providecommand \href [0]{\begingroup \@sanitize@url \@href}%
\providecommand \@href[1]{\@@startlink{#1}\@@href}%
\providecommand \@@href[1]{\endgroup#1\@@endlink}%
\providecommand \@sanitize@url [0]{\catcode `\\12\catcode `\$12\catcode
  `\&12\catcode `\#12\catcode `\^12\catcode `\_12\catcode `\%12\relax}%
\providecommand \@@startlink[1]{}%
\providecommand \@@endlink[0]{}%
\providecommand \url  [0]{\begingroup\@sanitize@url \@url }%
\providecommand \@url [1]{\endgroup\@href {#1}{\urlprefix }}%
\providecommand \urlprefix  [0]{URL }%
\providecommand \Eprint [0]{\href }%
\providecommand \doibase [0]{http://dx.doi.org/}%
\providecommand \selectlanguage [0]{\@gobble}%
\providecommand \bibinfo  [0]{\@secondoftwo}%
\providecommand \bibfield  [0]{\@secondoftwo}%
\providecommand \translation [1]{[#1]}%
\providecommand \BibitemOpen [0]{}%
\providecommand \bibitemStop [0]{}%
\providecommand \bibitemNoStop [0]{.\EOS\space}%
\providecommand \EOS [0]{\spacefactor3000\relax}%
\providecommand \BibitemShut  [1]{\csname bibitem#1\endcsname}%
\let\auto@bib@innerbib\@empty
\bibitem [{\citenamefont {Marpaung}\ \emph {et~al.}(2013)\citenamefont
  {Marpaung}, \citenamefont {Morrison}, \citenamefont {Pant},\ and\
  \citenamefont {Eggleton}}]{Marpaung2013}%
  \BibitemOpen
  \bibfield  {author} {\bibinfo {author} {\bibfnamefont {D.}~\bibnamefont
  {Marpaung}}, \bibinfo {author} {\bibfnamefont {B.}~\bibnamefont {Morrison}},
  \bibinfo {author} {\bibfnamefont {R.}~\bibnamefont {Pant}}, \ and\ \bibinfo
  {author} {\bibfnamefont {B.~J.}\ \bibnamefont {Eggleton}},\ }\href {\doibase
  10.1364/OL.38.004300} {\bibfield  {journal} {\bibinfo  {journal} {Optics
  Letters}\ }\textbf {\bibinfo {volume} {38}},\ \bibinfo {pages} {4300}
  (\bibinfo {year} {2013})}\BibitemShut {NoStop}%
\bibitem [{\citenamefont {Pagani}\ \emph {et~al.}(2014)\citenamefont {Pagani},
  \citenamefont {Marpaung}, \citenamefont {Choi}, \citenamefont {Madden},
  \citenamefont {Luther-Davies},\ and\ \citenamefont {Eggleton}}]{Pagani2014a}%
  \BibitemOpen
  \bibfield  {author} {\bibinfo {author} {\bibfnamefont {M.}~\bibnamefont
  {Pagani}}, \bibinfo {author} {\bibfnamefont {D.}~\bibnamefont {Marpaung}},
  \bibinfo {author} {\bibfnamefont {D.-Y.}\ \bibnamefont {Choi}}, \bibinfo
  {author} {\bibfnamefont {S.~J.}\ \bibnamefont {Madden}}, \bibinfo {author}
  {\bibfnamefont {B.}~\bibnamefont {Luther-Davies}}, \ and\ \bibinfo {author}
  {\bibfnamefont {B.~J.}\ \bibnamefont {Eggleton}},\ }\href {\doibase
  10.1364/OE.22.028810} {\bibfield  {journal} {\bibinfo  {journal} {Optics
  Express}\ }\textbf {\bibinfo {volume} {22}},\ \bibinfo {pages} {28810}
  (\bibinfo {year} {2014})}\BibitemShut {NoStop}%
\bibitem [{\citenamefont {Santagiustina}\ \emph {et~al.}(2013)\citenamefont
  {Santagiustina}, \citenamefont {Chin}, \citenamefont {Primerov},
  \citenamefont {Ursini},\ and\ \citenamefont
  {Th{\'{e}}venaz}}]{Santagiustina2013}%
  \BibitemOpen
  \bibfield  {author} {\bibinfo {author} {\bibfnamefont {M.}~\bibnamefont
  {Santagiustina}}, \bibinfo {author} {\bibfnamefont {S.}~\bibnamefont {Chin}},
  \bibinfo {author} {\bibfnamefont {N.}~\bibnamefont {Primerov}}, \bibinfo
  {author} {\bibfnamefont {L.}~\bibnamefont {Ursini}}, \ and\ \bibinfo {author}
  {\bibfnamefont {L.}~\bibnamefont {Th{\'{e}}venaz}},\ }\href {\doibase
  10.1038/srep01594} {\bibfield  {journal} {\bibinfo  {journal} {Scientific
  Reports}\ }\textbf {\bibinfo {volume} {3}},\ \bibinfo {pages} {1594}
  (\bibinfo {year} {2013})}\BibitemShut {NoStop}%
\bibitem [{\citenamefont {Li}\ \emph {et~al.}(2013)\citenamefont {Li},
  \citenamefont {Lee},\ and\ \citenamefont {Vahala}}]{Li2013}%
  \BibitemOpen
  \bibfield  {author} {\bibinfo {author} {\bibfnamefont {J.}~\bibnamefont
  {Li}}, \bibinfo {author} {\bibfnamefont {H.}~\bibnamefont {Lee}}, \ and\
  \bibinfo {author} {\bibfnamefont {K.~J.}\ \bibnamefont {Vahala}},\ }\href
  {http://www.nature.com/doifinder/10.1038/ncomms3097} {\bibfield  {journal}
  {\bibinfo  {journal} {Nature Communications}\ }\textbf {\bibinfo {volume}
  {4}} (\bibinfo {year} {2013})}\BibitemShut {NoStop}%
\bibitem [{\citenamefont {Merklein}\ \emph
  {et~al.}(2016{\natexlab{a}})\citenamefont {Merklein}, \citenamefont
  {Stiller}, \citenamefont {Kabakova}, \citenamefont {Mutugala}, \citenamefont
  {Vu}, \citenamefont {Madden}, \citenamefont {Eggleton},\ and\ \citenamefont
  {Slav{\'{i}}k}}]{Merklein2016}%
  \BibitemOpen
  \bibfield  {author} {\bibinfo {author} {\bibfnamefont {M.}~\bibnamefont
  {Merklein}}, \bibinfo {author} {\bibfnamefont {B.}~\bibnamefont {Stiller}},
  \bibinfo {author} {\bibfnamefont {I.~V.}\ \bibnamefont {Kabakova}}, \bibinfo
  {author} {\bibfnamefont {U.~S.}\ \bibnamefont {Mutugala}}, \bibinfo {author}
  {\bibfnamefont {K.}~\bibnamefont {Vu}}, \bibinfo {author} {\bibfnamefont
  {S.~J.}\ \bibnamefont {Madden}}, \bibinfo {author} {\bibfnamefont {B.~J.}\
  \bibnamefont {Eggleton}}, \ and\ \bibinfo {author} {\bibfnamefont
  {R.}~\bibnamefont {Slav{\'{i}}k}},\ }\href {\doibase 10.1364/OL.41.004633}
  {\bibfield  {journal} {\bibinfo  {journal} {Optics Letters}\ }\textbf
  {\bibinfo {volume} {41}},\ \bibinfo {pages} {4633} (\bibinfo {year}
  {2016}{\natexlab{a}})}\BibitemShut {NoStop}%
\bibitem [{\citenamefont {Merklein}\ \emph
  {et~al.}(2016{\natexlab{b}})\citenamefont {Merklein}, \citenamefont
  {Stiller}, \citenamefont {Vu}, \citenamefont {Madden},\ and\ \citenamefont
  {Eggleton}}]{Merklein2017}%
  \BibitemOpen
  \bibfield  {author} {\bibinfo {author} {\bibfnamefont {M.}~\bibnamefont
  {Merklein}}, \bibinfo {author} {\bibfnamefont {B.}~\bibnamefont {Stiller}},
  \bibinfo {author} {\bibfnamefont {K.}~\bibnamefont {Vu}}, \bibinfo {author}
  {\bibfnamefont {S.~J.}\ \bibnamefont {Madden}}, \ and\ \bibinfo {author}
  {\bibfnamefont {B.~J.}\ \bibnamefont {Eggleton}},\ }\href {\doibase
  10.1038/s41467-017-00717-y} {\bibfield  {journal} {\bibinfo  {journal}
  {Nature Communications}\ }\textbf {\bibinfo {volume} {8}},\ \bibinfo {pages}
  {574} (\bibinfo {year} {2016}{\natexlab{b}})},\ \Eprint
  {http://arxiv.org/abs/1608.08767} {arXiv:1608.08767} \BibitemShut {NoStop}%
\bibitem [{\citenamefont {Dong}\ \emph {et~al.}(2015)\citenamefont {Dong},
  \citenamefont {Shen}, \citenamefont {Zou}, \citenamefont {Zhang},
  \citenamefont {Fu},\ and\ \citenamefont {Guo}}]{Dong2015}%
  \BibitemOpen
  \bibfield  {author} {\bibinfo {author} {\bibfnamefont {C.-H.}\ \bibnamefont
  {Dong}}, \bibinfo {author} {\bibfnamefont {Z.}~\bibnamefont {Shen}}, \bibinfo
  {author} {\bibfnamefont {C.-L.}\ \bibnamefont {Zou}}, \bibinfo {author}
  {\bibfnamefont {Y.-L.}\ \bibnamefont {Zhang}}, \bibinfo {author}
  {\bibfnamefont {W.}~\bibnamefont {Fu}}, \ and\ \bibinfo {author}
  {\bibfnamefont {G.-C.}\ \bibnamefont {Guo}},\ }\href {\doibase
  10.1038/ncomms7193} {\bibfield  {journal} {\bibinfo  {journal} {Nature
  Communications}\ }\textbf {\bibinfo {volume} {6}},\ \bibinfo {pages} {6193}
  (\bibinfo {year} {2015})},\ \Eprint {http://arxiv.org/abs/1408.2606}
  {arXiv:1408.2606} \BibitemShut {NoStop}%
\bibitem [{\citenamefont {Galland}\ \emph {et~al.}(2014)\citenamefont
  {Galland}, \citenamefont {Sangouard}, \citenamefont {Piro}, \citenamefont
  {Gisin},\ and\ \citenamefont {Kippenberg}}]{Galland2014}%
  \BibitemOpen
  \bibfield  {author} {\bibinfo {author} {\bibfnamefont {C.}~\bibnamefont
  {Galland}}, \bibinfo {author} {\bibfnamefont {N.}~\bibnamefont {Sangouard}},
  \bibinfo {author} {\bibfnamefont {N.}~\bibnamefont {Piro}}, \bibinfo {author}
  {\bibfnamefont {N.}~\bibnamefont {Gisin}}, \ and\ \bibinfo {author}
  {\bibfnamefont {T.~J.}\ \bibnamefont {Kippenberg}},\ }\href {\doibase
  10.1103/PhysRevLett.112.143602} {\bibfield  {journal} {\bibinfo  {journal}
  {Physical Review Letters}\ }\textbf {\bibinfo {volume} {112}},\ \bibinfo
  {pages} {1} (\bibinfo {year} {2014})},\ \Eprint
  {http://arxiv.org/abs/1312.4303} {arXiv:1312.4303} \BibitemShut {NoStop}%
\bibitem [{\citenamefont {Fang}\ \emph {et~al.}(2016)\citenamefont {Fang},
  \citenamefont {Matheny}, \citenamefont {Luan},\ and\ \citenamefont
  {Painter}}]{Fang2016b}%
  \BibitemOpen
  \bibfield  {author} {\bibinfo {author} {\bibfnamefont {K.}~\bibnamefont
  {Fang}}, \bibinfo {author} {\bibfnamefont {M.~H.}\ \bibnamefont {Matheny}},
  \bibinfo {author} {\bibfnamefont {X.}~\bibnamefont {Luan}}, \ and\ \bibinfo
  {author} {\bibfnamefont {O.}~\bibnamefont {Painter}},\ }\href {\doibase
  10.1038/nphoton.2016.107} {\bibfield  {journal} {\bibinfo  {journal} {Nature
  Photonics}\ }\textbf {\bibinfo {volume} {10}},\ \bibinfo {pages} {489}
  (\bibinfo {year} {2016})}\BibitemShut {NoStop}%
\bibitem [{\citenamefont {Hotate}\ \emph {et~al.}(2012)\citenamefont {Hotate},
  \citenamefont {Watanabe}, \citenamefont {He},\ and\ \citenamefont
  {Kishi}}]{Hotate2012}%
  \BibitemOpen
  \bibfield  {author} {\bibinfo {author} {\bibfnamefont {K.}~\bibnamefont
  {Hotate}}, \bibinfo {author} {\bibfnamefont {R.}~\bibnamefont {Watanabe}},
  \bibinfo {author} {\bibfnamefont {Z.}~\bibnamefont {He}}, \ and\ \bibinfo
  {author} {\bibfnamefont {M.}~\bibnamefont {Kishi}},\ }\href@noop {}
  {\bibfield  {journal} {\bibinfo  {journal} {22nd International Conference on
  Optical Fiber Sensors}\ }\textbf {\bibinfo {volume} {8421}},\ \bibinfo
  {pages} {8421CE} (\bibinfo {year} {2012})}\BibitemShut {NoStop}%
\bibitem [{\citenamefont {Song}\ \emph {et~al.}(2006)\citenamefont {Song},
  \citenamefont {He},\ and\ \citenamefont {Hotate}}]{Song2006}%
  \BibitemOpen
  \bibfield  {author} {\bibinfo {author} {\bibfnamefont {K.~Y.}\ \bibnamefont
  {Song}}, \bibinfo {author} {\bibfnamefont {Z.}~\bibnamefont {He}}, \ and\
  \bibinfo {author} {\bibfnamefont {K.}~\bibnamefont {Hotate}},\ }\href
  {\doibase 10.1364/OL.31.002526} {\bibfield  {journal} {\bibinfo  {journal}
  {Optics Letters}\ }\textbf {\bibinfo {volume} {31}},\ \bibinfo {pages} {2526}
  (\bibinfo {year} {2006})}\BibitemShut {NoStop}%
\bibitem [{\citenamefont {Antman}\ \emph {et~al.}(2016)\citenamefont {Antman},
  \citenamefont {Clain}, \citenamefont {London},\ and\ \citenamefont
  {Zadok}}]{Antman2016}%
  \BibitemOpen
  \bibfield  {author} {\bibinfo {author} {\bibfnamefont {Y.}~\bibnamefont
  {Antman}}, \bibinfo {author} {\bibfnamefont {A.}~\bibnamefont {Clain}},
  \bibinfo {author} {\bibfnamefont {Y.}~\bibnamefont {London}}, \ and\ \bibinfo
  {author} {\bibfnamefont {A.}~\bibnamefont {Zadok}},\ }\href {\doibase
  10.1364/OPTICA.3.000510} {\bibfield  {journal} {\bibinfo  {journal} {Optica}\
  }\textbf {\bibinfo {volume} {3}},\ \bibinfo {pages} {510} (\bibinfo {year}
  {2016})}\BibitemShut {NoStop}%
\bibitem [{\citenamefont {Jia}\ \emph {et~al.}(2017)\citenamefont {Jia},
  \citenamefont {Chang}, \citenamefont {Lin}, \citenamefont {Xu},\ and\
  \citenamefont {Wu}}]{Jia2017}%
  \BibitemOpen
  \bibfield  {author} {\bibinfo {author} {\bibfnamefont {X.-H.}\ \bibnamefont
  {Jia}}, \bibinfo {author} {\bibfnamefont {H.-Q.}\ \bibnamefont {Chang}},
  \bibinfo {author} {\bibfnamefont {K.}~\bibnamefont {Lin}}, \bibinfo {author}
  {\bibfnamefont {C.}~\bibnamefont {Xu}}, \ and\ \bibinfo {author}
  {\bibfnamefont {J.-G.}\ \bibnamefont {Wu}},\ }\href {\doibase
  10.1364/OE.25.006997} {\bibfield  {journal} {\bibinfo  {journal} {Optics
  Express}\ }\textbf {\bibinfo {volume} {25}},\ \bibinfo {pages} {6997}
  (\bibinfo {year} {2017})}\BibitemShut {NoStop}%
\bibitem [{\citenamefont {Godet}\ \emph {et~al.}(2017)\citenamefont {Godet},
  \citenamefont {Ndao}, \citenamefont {Sylvestre}, \citenamefont {Pecheur},
  \citenamefont {Lebrun}, \citenamefont {Pauliat}, \citenamefont {Beugnot},\
  and\ \citenamefont {{Phan Huy}}}]{Godet2017}%
  \BibitemOpen
  \bibfield  {author} {\bibinfo {author} {\bibfnamefont {A.}~\bibnamefont
  {Godet}}, \bibinfo {author} {\bibfnamefont {A.}~\bibnamefont {Ndao}},
  \bibinfo {author} {\bibfnamefont {T.}~\bibnamefont {Sylvestre}}, \bibinfo
  {author} {\bibfnamefont {V.}~\bibnamefont {Pecheur}}, \bibinfo {author}
  {\bibfnamefont {S.}~\bibnamefont {Lebrun}}, \bibinfo {author} {\bibfnamefont
  {G.}~\bibnamefont {Pauliat}}, \bibinfo {author} {\bibfnamefont {J.-C.}\
  \bibnamefont {Beugnot}}, \ and\ \bibinfo {author} {\bibfnamefont
  {K.}~\bibnamefont {{Phan Huy}}},\ }\href {\doibase 10.1364/OPTICA.4.001232}
  {\bibfield  {journal} {\bibinfo  {journal} {Optica}\ }\textbf {\bibinfo
  {volume} {4}},\ \bibinfo {pages} {1232} (\bibinfo {year} {2017})},\ \Eprint
  {http://arxiv.org/abs/1706.03990} {arXiv:1706.03990} \BibitemShut {NoStop}%
\bibitem [{\citenamefont {Nikl{\`{e}}s}\ \emph {et~al.}(1996)\citenamefont
  {Nikl{\`{e}}s}, \citenamefont {Th{\'{e}}venaz},\ and\ \citenamefont
  {Robert}}]{Nikles1996a}%
  \BibitemOpen
  \bibfield  {author} {\bibinfo {author} {\bibfnamefont {M.}~\bibnamefont
  {Nikl{\`{e}}s}}, \bibinfo {author} {\bibfnamefont {L.}~\bibnamefont
  {Th{\'{e}}venaz}}, \ and\ \bibinfo {author} {\bibfnamefont {P.~A.}\
  \bibnamefont {Robert}},\ }\href {\doibase 10.1364/OL.21.000758} {\bibfield
  {journal} {\bibinfo  {journal} {Optics Letters}\ }\textbf {\bibinfo {volume}
  {21}},\ \bibinfo {pages} {758} (\bibinfo {year} {1996})}\BibitemShut
  {NoStop}%
\bibitem [{\citenamefont {Bao}\ \emph {et~al.}(1993)\citenamefont {Bao},
  \citenamefont {Webb},\ and\ \citenamefont {Jackson}}]{Bao1993a}%
  \BibitemOpen
  \bibfield  {author} {\bibinfo {author} {\bibfnamefont {X.}~\bibnamefont
  {Bao}}, \bibinfo {author} {\bibfnamefont {D.~J.}\ \bibnamefont {Webb}}, \
  and\ \bibinfo {author} {\bibfnamefont {D.~A.}\ \bibnamefont {Jackson}},\
  }\href@noop {} {\bibfield  {journal} {\bibinfo  {journal} {Opt. Lett.}\
  }\textbf {\bibinfo {volume} {18}},\ \bibinfo {pages} {1561} (\bibinfo {year}
  {1993})}\BibitemShut {NoStop}%
\bibitem [{\citenamefont {Chrostowski}\ and\ \citenamefont
  {Hochberg}(2015)}]{Chrostowski2013}%
  \BibitemOpen
  \bibfield  {author} {\bibinfo {author} {\bibfnamefont {L.}~\bibnamefont
  {Chrostowski}}\ and\ \bibinfo {author} {\bibfnamefont {M.}~\bibnamefont
  {Hochberg}},\ }\href {\doibase 10.1017/CBO9781316084168} {\emph {\bibinfo
  {title} {{Silicon Photonics Design}}}}\ (\bibinfo  {publisher} {Cambridge
  University Press},\ \bibinfo {address} {Cambridge},\ \bibinfo {year}
  {2015})\BibitemShut {NoStop}%
\bibitem [{\citenamefont {Poulton}\ \emph {et~al.}(2013)\citenamefont
  {Poulton}, \citenamefont {Pant},\ and\ \citenamefont
  {Eggleton}}]{Poulton2013}%
  \BibitemOpen
  \bibfield  {author} {\bibinfo {author} {\bibfnamefont {C.~G.}\ \bibnamefont
  {Poulton}}, \bibinfo {author} {\bibfnamefont {R.}~\bibnamefont {Pant}}, \
  and\ \bibinfo {author} {\bibfnamefont {B.~J.}\ \bibnamefont {Eggleton}},\
  }\href {\doibase 10.1364/JOSAB.30.002657} {\bibfield  {journal} {\bibinfo
  {journal} {Journal of the Optical Society of America B}\ }\textbf {\bibinfo
  {volume} {30}},\ \bibinfo {pages} {2657} (\bibinfo {year}
  {2013})}\BibitemShut {NoStop}%
\bibitem [{\citenamefont {Kittlaus}\ \emph {et~al.}(2016)\citenamefont
  {Kittlaus}, \citenamefont {Shin},\ and\ \citenamefont
  {Rakich}}]{Kittlaus2015}%
  \BibitemOpen
  \bibfield  {author} {\bibinfo {author} {\bibfnamefont {E.~A.}\ \bibnamefont
  {Kittlaus}}, \bibinfo {author} {\bibfnamefont {H.}~\bibnamefont {Shin}}, \
  and\ \bibinfo {author} {\bibfnamefont {P.~T.}\ \bibnamefont {Rakich}},\
  }\href {\doibase 10.1038/nphoton.2016.112} {\bibfield  {journal} {\bibinfo
  {journal} {Nature Photonics}\ }\textbf {\bibinfo {volume} {10}},\ \bibinfo
  {pages} {463} (\bibinfo {year} {2016})}\BibitemShut {NoStop}%
\bibitem [{\citenamefont {{Van Laer}}\ \emph {et~al.}(2015)\citenamefont {{Van
  Laer}}, \citenamefont {Kuyken}, \citenamefont {{Van Thourhout}},\ and\
  \citenamefont {Baets}}]{VanLaer2015}%
  \BibitemOpen
  \bibfield  {author} {\bibinfo {author} {\bibfnamefont {R.}~\bibnamefont {{Van
  Laer}}}, \bibinfo {author} {\bibfnamefont {B.}~\bibnamefont {Kuyken}},
  \bibinfo {author} {\bibfnamefont {D.}~\bibnamefont {{Van Thourhout}}}, \ and\
  \bibinfo {author} {\bibfnamefont {R.}~\bibnamefont {Baets}},\ }\href
  {\doibase 10.1038/nphoton.2015.11} {\bibfield  {journal} {\bibinfo  {journal}
  {Nature Photonics}\ }\textbf {\bibinfo {volume} {9}},\ \bibinfo {pages} {199}
  (\bibinfo {year} {2015})},\ \Eprint {http://arxiv.org/abs/1407.4977}
  {arXiv:1407.4977} \BibitemShut {NoStop}%
\bibitem [{\citenamefont {Morrison}\ \emph {et~al.}(2017)\citenamefont
  {Morrison}, \citenamefont {Casas-Bedoya}, \citenamefont {Ren}, \citenamefont
  {Vu}, \citenamefont {Liu}, \citenamefont {Zarifi}, \citenamefont {Nguyen},
  \citenamefont {Choi}, \citenamefont {Marpaung}, \citenamefont {Madden},
  \citenamefont {Mitchell},\ and\ \citenamefont {Eggleton}}]{Morrison2017}%
  \BibitemOpen
  \bibfield  {author} {\bibinfo {author} {\bibfnamefont {B.}~\bibnamefont
  {Morrison}}, \bibinfo {author} {\bibfnamefont {A.}~\bibnamefont
  {Casas-Bedoya}}, \bibinfo {author} {\bibfnamefont {G.}~\bibnamefont {Ren}},
  \bibinfo {author} {\bibfnamefont {K.}~\bibnamefont {Vu}}, \bibinfo {author}
  {\bibfnamefont {Y.}~\bibnamefont {Liu}}, \bibinfo {author} {\bibfnamefont
  {A.}~\bibnamefont {Zarifi}}, \bibinfo {author} {\bibfnamefont {T.~G.}\
  \bibnamefont {Nguyen}}, \bibinfo {author} {\bibfnamefont {D.-Y.}\
  \bibnamefont {Choi}}, \bibinfo {author} {\bibfnamefont {D.}~\bibnamefont
  {Marpaung}}, \bibinfo {author} {\bibfnamefont {S.~J.}\ \bibnamefont
  {Madden}}, \bibinfo {author} {\bibfnamefont {A.}~\bibnamefont {Mitchell}}, \
  and\ \bibinfo {author} {\bibfnamefont {B.~J.}\ \bibnamefont {Eggleton}},\
  }\href {\doibase 10.1364/OPTICA.4.000847} {\bibfield  {journal} {\bibinfo
  {journal} {Optica}\ }\textbf {\bibinfo {volume} {4}},\ \bibinfo {pages} {847}
  (\bibinfo {year} {2017})},\ \Eprint {http://arxiv.org/abs/1702.05233}
  {arXiv:1702.05233} \BibitemShut {NoStop}%
\bibitem [{\citenamefont {Wolff}\ \emph {et~al.}(2016)\citenamefont {Wolff},
  \citenamefont {Laer}, \citenamefont {Steel}, \citenamefont {Eggleton},\ and\
  \citenamefont {Poulton}}]{Wolff2016}%
  \BibitemOpen
  \bibfield  {author} {\bibinfo {author} {\bibfnamefont {C.}~\bibnamefont
  {Wolff}}, \bibinfo {author} {\bibfnamefont {R.~V.}\ \bibnamefont {Laer}},
  \bibinfo {author} {\bibfnamefont {M.~J.}\ \bibnamefont {Steel}}, \bibinfo
  {author} {\bibfnamefont {B.~J.}\ \bibnamefont {Eggleton}}, \ and\ \bibinfo
  {author} {\bibfnamefont {C.~G.}\ \bibnamefont {Poulton}},\ }\href {\doibase
  10.1088/1367-2630/18/2/025006} {\bibfield  {journal} {\bibinfo  {journal}
  {New Journal of Physics}\ }\textbf {\bibinfo {volume} {18}},\ \bibinfo
  {pages} {025006} (\bibinfo {year} {2016})},\ \Eprint
  {http://arxiv.org/abs/1510.00079} {arXiv:1510.00079} \BibitemShut {NoStop}%
\bibitem [{\citenamefont {Chow}\ \emph {et~al.}(2018)\citenamefont {Chow},
  \citenamefont {Beugnot}, \citenamefont {Godet}, \citenamefont {Huy},
  \citenamefont {Soto},\ and\ \citenamefont {Th{\'{e}}venaz}}]{Chow2018}%
  \BibitemOpen
  \bibfield  {author} {\bibinfo {author} {\bibfnamefont {D.~M.}\ \bibnamefont
  {Chow}}, \bibinfo {author} {\bibfnamefont {J.-C.}\ \bibnamefont {Beugnot}},
  \bibinfo {author} {\bibfnamefont {A.}~\bibnamefont {Godet}}, \bibinfo
  {author} {\bibfnamefont {K.~P.}\ \bibnamefont {Huy}}, \bibinfo {author}
  {\bibfnamefont {M.~A.}\ \bibnamefont {Soto}}, \ and\ \bibinfo {author}
  {\bibfnamefont {L.}~\bibnamefont {Th{\'{e}}venaz}},\ }\href {\doibase
  10.1364/OL.43.001487} {\bibfield  {journal} {\bibinfo  {journal} {Optics
  Letters}\ }\textbf {\bibinfo {volume} {43}},\ \bibinfo {pages} {1487}
  (\bibinfo {year} {2018})}\BibitemShut {NoStop}%
\bibitem [{\citenamefont {Stiller}\ \emph {et~al.}(2010)\citenamefont
  {Stiller}, \citenamefont {Foaleng}, \citenamefont {Beugnot}, \citenamefont
  {Lee}, \citenamefont {Delqu{\'{e}}}, \citenamefont {Bouwmans}, \citenamefont
  {Kudlinski}, \citenamefont {Th{\'{e}}venaz}, \citenamefont {Maillotte},\ and\
  \citenamefont {Sylvestre}}]{Stiller2010}%
  \BibitemOpen
  \bibfield  {author} {\bibinfo {author} {\bibfnamefont {B.}~\bibnamefont
  {Stiller}}, \bibinfo {author} {\bibfnamefont {S.~M.}\ \bibnamefont
  {Foaleng}}, \bibinfo {author} {\bibfnamefont {J.-C.}\ \bibnamefont
  {Beugnot}}, \bibinfo {author} {\bibfnamefont {M.~W.}\ \bibnamefont {Lee}},
  \bibinfo {author} {\bibfnamefont {M.}~\bibnamefont {Delqu{\'{e}}}}, \bibinfo
  {author} {\bibfnamefont {G.}~\bibnamefont {Bouwmans}}, \bibinfo {author}
  {\bibfnamefont {A.}~\bibnamefont {Kudlinski}}, \bibinfo {author}
  {\bibfnamefont {L.}~\bibnamefont {Th{\'{e}}venaz}}, \bibinfo {author}
  {\bibfnamefont {H.}~\bibnamefont {Maillotte}}, \ and\ \bibinfo {author}
  {\bibfnamefont {T.}~\bibnamefont {Sylvestre}},\ }\href {\doibase
  10.1364/OE.18.020136} {\bibfield  {journal} {\bibinfo  {journal} {Optics
  Express}\ }\textbf {\bibinfo {volume} {18}},\ \bibinfo {pages} {20136}
  (\bibinfo {year} {2010})}\BibitemShut {NoStop}%
\bibitem [{\citenamefont {Stiller}\ \emph {et~al.}(2012)\citenamefont
  {Stiller}, \citenamefont {Kudlinski}, \citenamefont {Lee}, \citenamefont
  {Bouwmans}, \citenamefont {Delque}, \citenamefont {Beugnot}, \citenamefont
  {Maillotte},\ and\ \citenamefont {Sylvestre}}]{B.Stiller2012}%
  \BibitemOpen
  \bibfield  {author} {\bibinfo {author} {\bibfnamefont {B.}~\bibnamefont
  {Stiller}}, \bibinfo {author} {\bibfnamefont {A.}~\bibnamefont {Kudlinski}},
  \bibinfo {author} {\bibfnamefont {M.~W.}\ \bibnamefont {Lee}}, \bibinfo
  {author} {\bibfnamefont {G.}~\bibnamefont {Bouwmans}}, \bibinfo {author}
  {\bibfnamefont {M.}~\bibnamefont {Delque}}, \bibinfo {author} {\bibfnamefont
  {J.-C.}\ \bibnamefont {Beugnot}}, \bibinfo {author} {\bibfnamefont
  {H.}~\bibnamefont {Maillotte}}, \ and\ \bibinfo {author} {\bibfnamefont
  {T.}~\bibnamefont {Sylvestre}},\ }\href {\doibase 10.1109/LPT.2012.2186286}
  {\bibfield  {journal} {\bibinfo  {journal} {IEEE Photonics Technology
  Letters}\ }\textbf {\bibinfo {volume} {24}},\ \bibinfo {pages} {667}
  (\bibinfo {year} {2012})}\BibitemShut {NoStop}%
\bibitem [{\citenamefont {Beugnot}\ \emph
  {et~al.}(2007{\natexlab{a}})\citenamefont {Beugnot}, \citenamefont
  {Sylvestre}, \citenamefont {Alasia}, \citenamefont {Maillotte}, \citenamefont
  {Laude}, \citenamefont {Monteville}, \citenamefont {Provino}, \citenamefont
  {Traynor}, \citenamefont {Mafang},\ and\ \citenamefont
  {Th{\'{e}}venaz}}]{Beugnot2007}%
  \BibitemOpen
  \bibfield  {author} {\bibinfo {author} {\bibfnamefont {J.-C.}\ \bibnamefont
  {Beugnot}}, \bibinfo {author} {\bibfnamefont {T.}~\bibnamefont {Sylvestre}},
  \bibinfo {author} {\bibfnamefont {D.}~\bibnamefont {Alasia}}, \bibinfo
  {author} {\bibfnamefont {H.}~\bibnamefont {Maillotte}}, \bibinfo {author}
  {\bibfnamefont {V.}~\bibnamefont {Laude}}, \bibinfo {author} {\bibfnamefont
  {A.}~\bibnamefont {Monteville}}, \bibinfo {author} {\bibfnamefont
  {L.}~\bibnamefont {Provino}}, \bibinfo {author} {\bibfnamefont
  {N.}~\bibnamefont {Traynor}}, \bibinfo {author} {\bibfnamefont {S.~F.}\
  \bibnamefont {Mafang}}, \ and\ \bibinfo {author} {\bibfnamefont
  {L.}~\bibnamefont {Th{\'{e}}venaz}},\ }\href {\doibase 10.1364/OE.15.015517}
  {\bibfield  {journal} {\bibinfo  {journal} {Optics Express}\ }\textbf
  {\bibinfo {volume} {15}},\ \bibinfo {pages} {15517} (\bibinfo {year}
  {2007}{\natexlab{a}})}\BibitemShut {NoStop}%
\bibitem [{\citenamefont {Zarifi}\ \emph {et~al.}(2018)\citenamefont {Zarifi},
  \citenamefont {Stiller}, \citenamefont {Merklein}, \citenamefont {Li},
  \citenamefont {Vu}, \citenamefont {Choi}, \citenamefont {Ma}, \citenamefont
  {Madden},\ and\ \citenamefont {Eggleton}}]{Zarifi2017}%
  \BibitemOpen
  \bibfield  {author} {\bibinfo {author} {\bibfnamefont {A.}~\bibnamefont
  {Zarifi}}, \bibinfo {author} {\bibfnamefont {B.}~\bibnamefont {Stiller}},
  \bibinfo {author} {\bibfnamefont {M.}~\bibnamefont {Merklein}}, \bibinfo
  {author} {\bibfnamefont {N.}~\bibnamefont {Li}}, \bibinfo {author}
  {\bibfnamefont {K.}~\bibnamefont {Vu}}, \bibinfo {author} {\bibfnamefont
  {D.-Y.}\ \bibnamefont {Choi}}, \bibinfo {author} {\bibfnamefont
  {P.}~\bibnamefont {Ma}}, \bibinfo {author} {\bibfnamefont {S.~J.}\
  \bibnamefont {Madden}}, \ and\ \bibinfo {author} {\bibfnamefont {B.~J.}\
  \bibnamefont {Eggleton}},\ }\href {\doibase 10.1063/1.5000108} {\bibfield
  {journal} {\bibinfo  {journal} {APL Photonics}\ }\textbf {\bibinfo {volume}
  {3}},\ \bibinfo {pages} {036101} (\bibinfo {year} {2018})},\ \Eprint
  {http://arxiv.org/abs/1707.09684} {arXiv:1707.09684} \BibitemShut {NoStop}%
\bibitem [{\citenamefont {Jarschel}\ \emph {et~al.}(2018)\citenamefont
  {Jarschel}, \citenamefont {Magalhaes}, \citenamefont {Aldaya}, \citenamefont
  {Florez},\ and\ \citenamefont {Dainese}}]{Jarschel2018}%
  \BibitemOpen
  \bibfield  {author} {\bibinfo {author} {\bibfnamefont {P.~F.}\ \bibnamefont
  {Jarschel}}, \bibinfo {author} {\bibfnamefont {L.~S.}\ \bibnamefont
  {Magalhaes}}, \bibinfo {author} {\bibfnamefont {I.}~\bibnamefont {Aldaya}},
  \bibinfo {author} {\bibfnamefont {O.}~\bibnamefont {Florez}}, \ and\ \bibinfo
  {author} {\bibfnamefont {P.}~\bibnamefont {Dainese}},\ }\href {\doibase
  10.1364/OL.43.000995} {\bibfield  {journal} {\bibinfo  {journal} {Optics
  Letters}\ }\textbf {\bibinfo {volume} {43}},\ \bibinfo {pages} {995}
  (\bibinfo {year} {2018})}\BibitemShut {NoStop}%
\bibitem [{\citenamefont {Florez}\ \emph {et~al.}(2016)\citenamefont {Florez},
  \citenamefont {Jarschel}, \citenamefont {Espinel}, \citenamefont {Cordeiro},
  \citenamefont {{Mayer Alegre}}, \citenamefont {Wiederhecker},\ and\
  \citenamefont {Dainese}}]{Florez2016a}%
  \BibitemOpen
  \bibfield  {author} {\bibinfo {author} {\bibfnamefont {O.}~\bibnamefont
  {Florez}}, \bibinfo {author} {\bibfnamefont {P.~F.}\ \bibnamefont
  {Jarschel}}, \bibinfo {author} {\bibfnamefont {Y.~A.~V.}\ \bibnamefont
  {Espinel}}, \bibinfo {author} {\bibfnamefont {C.~M.~B.}\ \bibnamefont
  {Cordeiro}}, \bibinfo {author} {\bibfnamefont {T.~P.}\ \bibnamefont {{Mayer
  Alegre}}}, \bibinfo {author} {\bibfnamefont {G.~S.}\ \bibnamefont
  {Wiederhecker}}, \ and\ \bibinfo {author} {\bibfnamefont {P.}~\bibnamefont
  {Dainese}},\ }\href {\doibase 10.1038/ncomms11759} {\bibfield  {journal}
  {\bibinfo  {journal} {Nature Communications}\ }\textbf {\bibinfo {volume}
  {7}},\ \bibinfo {pages} {11759} (\bibinfo {year} {2016})},\ \Eprint
  {http://arxiv.org/abs/1601.05248} {arXiv:1601.05248} \BibitemShut {NoStop}%
\bibitem [{\citenamefont {Kang}\ \emph {et~al.}(2008)\citenamefont {Kang},
  \citenamefont {Brenn}, \citenamefont {Wiederhecker},\ and\ \citenamefont
  {Russell}}]{Kang2008}%
  \BibitemOpen
  \bibfield  {author} {\bibinfo {author} {\bibfnamefont {M.~S.}\ \bibnamefont
  {Kang}}, \bibinfo {author} {\bibfnamefont {A.}~\bibnamefont {Brenn}},
  \bibinfo {author} {\bibfnamefont {G.~S.}\ \bibnamefont {Wiederhecker}}, \
  and\ \bibinfo {author} {\bibfnamefont {P.~S.}\ \bibnamefont {Russell}},\
  }\href {\doibase 10.1063/1.2995863} {\bibfield  {journal} {\bibinfo
  {journal} {Applied Physics Letters}\ }\textbf {\bibinfo {volume} {93}},\
  \bibinfo {pages} {131110} (\bibinfo {year} {2008})}\BibitemShut {NoStop}%
\bibitem [{\citenamefont {Beugnot}\ \emph
  {et~al.}(2007{\natexlab{b}})\citenamefont {Beugnot}, \citenamefont
  {Sylvestre}, \citenamefont {Maillotte}, \citenamefont {M{\'{e}}lin},\ and\
  \citenamefont {Laude}}]{Zhong2015a}%
  \BibitemOpen
  \bibfield  {author} {\bibinfo {author} {\bibfnamefont {J.-C.}\ \bibnamefont
  {Beugnot}}, \bibinfo {author} {\bibfnamefont {T.}~\bibnamefont {Sylvestre}},
  \bibinfo {author} {\bibfnamefont {H.}~\bibnamefont {Maillotte}}, \bibinfo
  {author} {\bibfnamefont {G.}~\bibnamefont {M{\'{e}}lin}}, \ and\ \bibinfo
  {author} {\bibfnamefont {V.}~\bibnamefont {Laude}},\ }\href {\doibase
  10.1364/OL.32.000017} {\bibfield  {journal} {\bibinfo  {journal} {Optics
  Letters}\ }\textbf {\bibinfo {volume} {32}},\ \bibinfo {pages} {17} (\bibinfo
  {year} {2007}{\natexlab{b}})},\ \Eprint {http://arxiv.org/abs/1508.01309}
  {arXiv:1508.01309} \BibitemShut {NoStop}%
\bibitem [{\citenamefont {Cohen}\ \emph {et~al.}(2014)\citenamefont {Cohen},
  \citenamefont {London}, \citenamefont {Antman},\ and\ \citenamefont
  {Zadok}}]{Cohen2014a}%
  \BibitemOpen
  \bibfield  {author} {\bibinfo {author} {\bibfnamefont {R.}~\bibnamefont
  {Cohen}}, \bibinfo {author} {\bibfnamefont {Y.}~\bibnamefont {London}},
  \bibinfo {author} {\bibfnamefont {Y.}~\bibnamefont {Antman}}, \ and\ \bibinfo
  {author} {\bibfnamefont {A.}~\bibnamefont {Zadok}},\ }\href {\doibase
  10.1364/OE.22.012070} {\bibfield  {journal} {\bibinfo  {journal} {Optics
  Express}\ }\textbf {\bibinfo {volume} {22}},\ \bibinfo {pages} {12070}
  (\bibinfo {year} {2014})}\BibitemShut {NoStop}%
\bibitem [{\citenamefont {Eggleton}\ \emph {et~al.}(2013)\citenamefont
  {Eggleton}, \citenamefont {Poulton},\ and\ \citenamefont
  {Pant}}]{Eggleton2013}%
  \BibitemOpen
  \bibfield  {author} {\bibinfo {author} {\bibfnamefont {B.~J.}\ \bibnamefont
  {Eggleton}}, \bibinfo {author} {\bibfnamefont {C.~G.}\ \bibnamefont
  {Poulton}}, \ and\ \bibinfo {author} {\bibfnamefont {R.}~\bibnamefont
  {Pant}},\ }\href {\doibase 10.1364/AOP.5.000536} {\bibfield  {journal}
  {\bibinfo  {journal} {Advances in Optics and Photonics}\ }\textbf {\bibinfo
  {volume} {5}},\ \bibinfo {pages} {536} (\bibinfo {year} {2013})}\BibitemShut
  {NoStop}%
\bibitem [{\citenamefont {Fellay}\ \emph {et~al.}(1997)\citenamefont {Fellay},
  \citenamefont {Th{\'{e}}venaz}, \citenamefont {Facchini}, \citenamefont
  {Nikl{\`{e}}s},\ and\ \citenamefont {Robert}}]{Fellay1997}%
  \BibitemOpen
  \bibfield  {author} {\bibinfo {author} {\bibfnamefont {A.}~\bibnamefont
  {Fellay}}, \bibinfo {author} {\bibfnamefont {L.}~\bibnamefont
  {Th{\'{e}}venaz}}, \bibinfo {author} {\bibfnamefont {M.}~\bibnamefont
  {Facchini}}, \bibinfo {author} {\bibfnamefont {M.}~\bibnamefont
  {Nikl{\`{e}}s}}, \ and\ \bibinfo {author} {\bibfnamefont {P.}~\bibnamefont
  {Robert}},\ }\href@noop {} {\bibfield  {journal} {\bibinfo  {journal} {12th
  International Conference on Optical Fiber Sensors}\ }\textbf {\bibinfo
  {volume} {16}} (\bibinfo {year} {1997})}\BibitemShut {NoStop}%
\end{thebibliography}%



\end{document}